\documentclass[prapplied,longbibliography,
superscriptaddress]{revtex4-1}
\usepackage{graphicx}
\usepackage{hyperref}
\usepackage{dcolumn}   
\usepackage{bm}        
\usepackage{amssymb}   
\usepackage[english]{babel} 
\usepackage{inputenc}
\usepackage{amsmath}
\usepackage{amssymb}
\usepackage{graphicx}
\usepackage{natbib}
\usepackage{mathrsfs}
\usepackage{color}
\usepackage{upgreek}

\begin{document}
\title{Nanomechanical Dissipation and Strain Engineering}

\author{Leo Sementilli}
\author{Erick Romero}
\author{Warwick P. Bowen}
\email{w.bowen@uq.edu.au}
\affiliation{The Australian Research Council Centre of Excellence for Engineered Quantum Systems, School of Mathematics and Physics, University of Queensland, St. Lucia, Queensland 4072, Australia}

\begin{abstract}

Nanomechanical resonators have applications in a wide variety of technologies ranging from biochemical sensors to mobile communications, quantum computing, inertial sensing, and precision navigation. The quality factor of the mechanical resonance is critical for many applications. Until recently, mechanical quality factors rarely exceeded a million. In the past few years however, new methods have been developed to exceed this boundary. These methods involve careful engineering of the structure of the nanomechanical resonator, including the use of acoustic bandgaps and nested structures to suppress dissipation into the substrate, and the use of dissipation dilution and strain engineering to increase the mechanical frequency and suppress intrinsic dissipation. Together, they have allowed quality factors to reach values near a billion at room temperature, resulting in exceptionally low dissipation.
This review aims to provide a pedagogical introduction to these new methods, primarily targeted to readers who are new to the field, together with an overview of the existing state-of-the-art, what may be possible in the future, and a perspective on the future applications of these extreme-high quality resonators.
\end{abstract}

\maketitle
\section{Introduction}

Nano- and micro-mechanical resonators have a broad range of applications, from micro-electromechanical sensors (MEMS)~\cite{bogue_recent_2013,villiers_electromechanical_2016}, to radio- and microwave frequency filters and clocks \cite{massel_microwave_2011,nguyen_mems_2007}, future computing technologies~\cite{pechal_superconducting_2018}, and tests of fundamental science \cite{forstner_nanomechanical_2020,vinante_improved_2017}. As sensors, they have been applied in inertial and gyroscopic navigation \cite{krause_high-resolution_2012,fan_graphene_2019,shaeffer_mems_2013,dabove_inertial_2015,cook_high-mass_2018,ayazi_harpss_2001,zhou_broadband_2021}, gas and biochemical sensing \cite{muruganathan_graphene_2018,eom_nanomechanical_2011,carrascosa_nanomechanical_2006}, atomic force microscopy, magnetometry~\cite{forstner2014ultrasensitive,li2020ultrabroadband,forstner2012cavity}, acoustic and temperature sensing~\cite{basiri-esfahani_precision_2019,firdous_assembling_2020,purdy2017quantum,singh2019detecting,westerveld_sensitive_2021}, single protein mass spectroscopy \cite{hanay_single-protein_2012} and single molecule detection \cite{chien_single-molecule_2018}. They are key elements for future computer technologies based on nanomechanical motion \cite{mauranyapin_tunneling_2021}, and for future quantum information technologies due to their ability to mediate various quantum degrees of freedom~\cite{clerk_hybrid_2020,hadfield_superconducting_2016,chu_quantum_2017,bowen_quantum_2016}. Recently, high quality factor nanomechanical resonators have enabled a broad range of new quantum physics, such as cooling of a nanomechanical oscillator to its quantum ground state~\cite{chan_laser_2011,teufel_sideband_2011}, entanglement with microwave fields~\cite{palomaki_entangling_2013,romero-sanchez_quantum_2018} and entanglement between mechanical resonators~\cite{ockeloen-korppi_stabilized_2018}.

In nearly all applications performance is dependent on the mechanical energy dissipation rate $\Gamma$, where a small dissipation rate is desirable to both isolate the resonator from environmental noise sources and to maintain coherent oscillation for an extended time. A more coherent resonator offers better stability and suppression of noise sources, thus allowing for cleaner and more precise readout \cite{walls_fundamental_1995,cleland_noise_2002,aspelmeyer_cavity_2014}.
The dissipation rate is often compared to the mechanical resonance frequency $\omega_\textnormal{m}$ via the 
mechanical quality factor $Q$~\cite{schmid_fundamentals_2016},
\begin{equation}\label{eq:Q_FreqDissipation}
Q= \frac{\omega_\textnormal{m}}{\Gamma}=2\pi \frac{W}{\Delta W}.
\end{equation}
The quality factor quantifies the number of coherent oscillations of the resonator before it decays to roughly 4\% percent of its original amplitude. As also expressed in Equation~(\ref{eq:Q_FreqDissipation}), the quality factor can equally be described as the quotient 
of stored energy in the resonator $W$ to the energy dissipated per cycle of oscillation $\Delta W/2\pi$.

Recent years have seen dramatic progress in our understanding of the fundamental origins of the primary dissipation mechanisms relevant to nanomechanical resonators. From this, new engineering techniques and strategies have emerged to suppress the dissipation. These new techniques include acoustic isolation \cite{tsaturyan_demonstration_2014,ghadimi_radiation_2017}, soft clamping \cite{tsaturyan_ultracoherent_2017,beccari_hierarchical_2021}, dissipation dilution \cite{fedorov_generalized_2019} and strain engineering \cite{ghadimi_elastic_2018}. Leveraging these techniques has enabled quality factors approaching one billion at room temperature in amorphous nanomechanical resonators~\cite{ghadimi_elastic_2018,tsaturyan_ultracoherent_2017,beccari_hierarchical_2021}, together with predictions that even higher quality factors should be possible using crystalline resonators \cite{romero_engineering_2020,ghadimi_elastic_2018}. This provides the prospect of orders-of-magnitude sensitivity enhancements in nanomechanical sensors~\cite{ekinci_ultimate_2004}, extremely narrow nanomechanical filters~\cite{sun_znosilicon_2012}, high density nanomechanical memories~\cite{merklein_chip-integrated_2017} and delay lines that can passively store information for many minutes~\cite{roodenburg_buckling_2009}, as well as fundamentally new technologies such as nanomechanical computer~\cite{pechal_superconducting_2018} and quantum information processors~\cite{kotler_direct_2021,mercier_de_lepinay_quantum_2021}. The purpose of this review is to provide an introduction to these new techniques, and an outlook on applications and what further advances may be possible in future.

\subsection{Fundamentals of a Nanomechanical String Resonator}\label{sec:funstring}

In this review, to provide concrete examples throughout, we consider the case of a string resonator under tensile stress. At the most basic level, string resonators are analogous to guitar strings. They feature two clamping points and are sufficiently thin such that their resonances have roughly sinusoidal eigenmodes (or modeshapes), for the most part unaffected by the presence of the clamping points. The fundamental mode of a string resonator has nodes only at the clamping points, and is labelled by the mode number $n=1$, whereas higher order modes have increasing numbers of nodes. An example of a string resonator is shown in Figure~\ref{fig:String} together with the shape of its first transversal modes. The mode shape $\phi(x,y,z)$ of a mechanical resonator defines a normalised amplitude of displacement of each spatial element of the resonator. For a string, it is generally assumed to be a scalar field, with displacement only occurring in the vertical direction, defined here as the $z$ direction. Moreover, for a sufficiently thin string resonator with uniform width $w$ the displacement is constant across its width \cite{fedorov_generalized_2019,ghadimi_ultra-coherent_2018}, here defined to be the $y$ direction. For simplicity we take these approximations for the majority of this review, so that the $n$-th string eigenmode can be simplified to a unidimensional problem $\phi_n(x)$, where $x$ is the longitudinal position along the string. This assumption is reasonable so long as the string has a uniform width and that its length $L$ is much larger than both its width $w$ and its thickness $h$ ($L \gg \{w,h\}$). 

Under the approximations outlined in the previous paragraph, the characteristic resonance frequencies, or  eigenfrequencies, of a string resonator of length $L$ and cross-section $A$ are given by \cite{schmid_fundamentals_2016}
\begin{equation}\label{eq:eigenfreq}
\omega_\textnormal{m}=\beta_{\sigma}^2 \sqrt{\frac{EI_y}{\rho A}}\sqrt{1+\frac{\sigma A}{E I_y \beta_{\sigma}^2}},
\end{equation}
where $\beta_{\sigma}=n\pi/L$ is the wavenumber for the mode $n$, $\sigma$ is the stress of the string, $I_y$ is the geometric moment of inertia, and $E$ and $\rho$ are the Young's modulus and density of the material respectively. Assuming a string with a rectangular cross-sectional area so that $A=w\times h$, the moment of inertia is given by $I_y=A h^2/12$.

\section{Mechanisms of elastic energy dissipation}
\label{mech_sec}

Micro- and nano-mechanical resonators are  simultaneously subjected to various damping mechanisms. The magnitude of each mechanisms contribution to the resonator's total dissipation typically depends on a large number of variables, often interconnected. These include the resonator's geometry, material, vibrational mode, resonance frequency, and surrounding pressure and temperature, among others. Identification and mitigation of these dissipation mechanisms in this multivariate space is a non-trivial task.

The dissipation mechanisms can be divided into \textit{intrinsic dissipation}, $\Gamma_{\textnormal{int}}$, internal to the resonator itself, and {\it extrinsic dissipation} (also termed {\it external dissipation}), $\Gamma_{\textnormal{ext}}$, related to the resonator's interactions with its environment. The total quality factor of the resonator can then be expressed as
\begin{equation}\label{eq:Q_extint}
\frac{1}{Q}= \frac{1}{Q_{\textnormal{int}}} +  \frac{1}{Q_{\textnormal{ext}}},
\end{equation}
where $Q_{\textnormal{int}}=\omega_\textnormal{m}/\Gamma_{\textnormal{int}}$ and $Q_{\textnormal{ext}}= \omega_\textnormal{m}/\Gamma_{\textnormal{ext}}$ are the intrinsic and extrinsic quality factors, respectively \cite{tsaturyan_ultracoherent_2019,ghadimi_ultra-coherent_2018}.
In the following we introduce the various forms of intrinsic and extrinsic dissipation mechanisms common to micro- and nano-mechanical resonators, their dependencies on system parameters, and how extrinsic mechanisms may be suppressed.

\subsection{Intrinsic Dissipation}\label{sec:IntrinsicLoss}

A resonator's intrinsic dissipation can be attributed to various processes occurring in the material's bulk and surface~\cite{villanueva_evidence_2014,mohanty_intrinsic_2002,jinling_yang_energy_2002}. This energy loss, inherent to the resonator, can be further divided into processes associated with the internal friction of the material and those more fundamental, unrelated to the frictional damping~\cite{schmid_fundamentals_2016}. The internal friction of a resonator is determined by the material type and quality, commonly more significant in amorphous and defect-rich materials~\cite{ashby_overview_1989}. As the resonator's atoms move during vibration, friction between them gives rise to damping and energy loss~\cite{zener_internal_1938}. This intrinsic friction can stem from defects, discontinuities in crystalline materials, or molecular chain movements in amorphous materials. Frictional loss is commonly modelled as a lagged response to a coherent excitation, given by the Zener's model of anelastic materials~\cite{zener_internal_1937,zener_internal_1938}. 

Independent of this frictional loss, additional sources of intrinsic dissipation exist in nanomechanical resonators. These loss processes arise from the coupling of the resonator's oscillating strain field to temperature fields or two-level systems~\cite{schmid_fundamentals_2016,tsaturyan_ultracoherent_2019}. Various causes of fundamental loss, such as thermoelastic damping, Akhiezer damping and loss to two-level systems, have been well studied~\cite{lifshitz_thermoelastic_2000,kunal_akhiezer_2011,ghaffari_quantum_2013,faust_signatures_2014,rodriguez_direct_2019,chakram_dissipation_2014}. For a more thorough explanation of the various intrinsic dissipation mechanisms, refer to Refs.~\cite{schmid_fundamentals_2016,joshi_design_2014,imboden_dissipation_2014}.

\subsection{Extrinsic Dissipation}\label{sec:ExtrinsicDissipation}

 Extrinsic dissipation can include gas damping, clamping loss, electrical charge damping and magnetomotive damping among other forms of dissipation~\cite{schmid_fundamentals_2016}. The total extrinsic quality factor is given in terms of the  quality factors $Q_i$ of these various damping mechanisms as
 \begin{equation}\label{eq:Q_ext}
Q_{\textnormal{ext}}^{-1}= \sum_{i} Q_{i}^{-1},
\end{equation} 
where the subscript $i$ labels the different mechanisms.

In applications where the resonator must interact with a surrounding medium, such as biochemical sensing~\cite{etayash_microfluidic_2016,fritz_translating_2000}, acoustic sensing~\cite{basiri-esfahani_precision_2019} and hydrodynamic experiments  \cite{ekinci_high-frequency_2010,
kara_nanofluidics_2015}, medium interaction damping is often the primary source of damping~\cite{naesby_effects_2017}.
As the name suggests, this loss arises from the interaction between a resonator and its surrounding medium. This medium, usually gas or liquid, damps the resonator's motion through the exchange of energy during molecular collisions or viscous interactions \cite{schmid_fundamentals_2016,sader_frequency_1998}.

Where the application does not depend on interactions with the surrounding medium -- such as inertial sensing~\cite{shaeffer_mems_2013,ekinci_ultimate_2004}, nanomechanical circuitry~\cite{hatanaka_phonon_2014,romero_propagation_2019} or quantum information applications~\cite{patel_single-mode_2018,pechal_superconducting_2018}, medium losses can be straightforwardly eliminated by placing the resonator in vacuum~\cite{naesby_effects_2017,verbridge_size_2008}. In this case, {\it clamping losses} are generally the dominant source of external loss.
Clamping losses, also known as {\it radiation loss}, arise from the modulation in strain at the clamping (or anchor) points at which the oscillating resonator is attached to its substrate.
 As a resonator oscillates it pulls on the clamping  points, inducing strain and subsequent acoustic waves \cite{photiadis_attachment_2004}. These acoustic waves radiate outward into the substrate carrying away energy that was previously stored in the resonator. The magnitude of clamping loss is dependent on both the impedance and mode matching between the resonator and substrate, as well as the geometry of the resonator \cite{schmid_fundamentals_2016,cross_elastic_2001}. Clamping loss has even been found to be sensitive to the way in which a chip is mounted, whether that be adhesive bonding, mechanical clamps or resting unclamped under gravity \cite{wilson_cavity_2009,schmid_damping_2011,wilson_cavity_nodate}.

\subsection{Suppressing Clamping Loss}\label{sec:SuppressExtrinsicLoss}

A range of techniques have been developed to suppress the radiation of acoustic energy at clamping points. Generally, they involve either increasing the impedance mismatch between the wave in the resonator and that in the substrate~\cite{norte_mechanical_2016,reinhardt_ultralow-noise_2016}, acoustic wave bandgaps~\cite{tsaturyan_demonstration_2014,yu_phononic_2014,ghadimi_radiation_2017,tsaturyan_ultracoherent_2017}, or acoustic filters based on nested structures \cite{weaver_nested_2016,romero_sanchez_phononics_2019,nielsen_multimode_2017}.

\subsubsection{Impedance Mismatch and Clamping Loss of a String Resonator}

\begin{figure}[h]
\begin{center}
\includegraphics[width=1\textwidth]{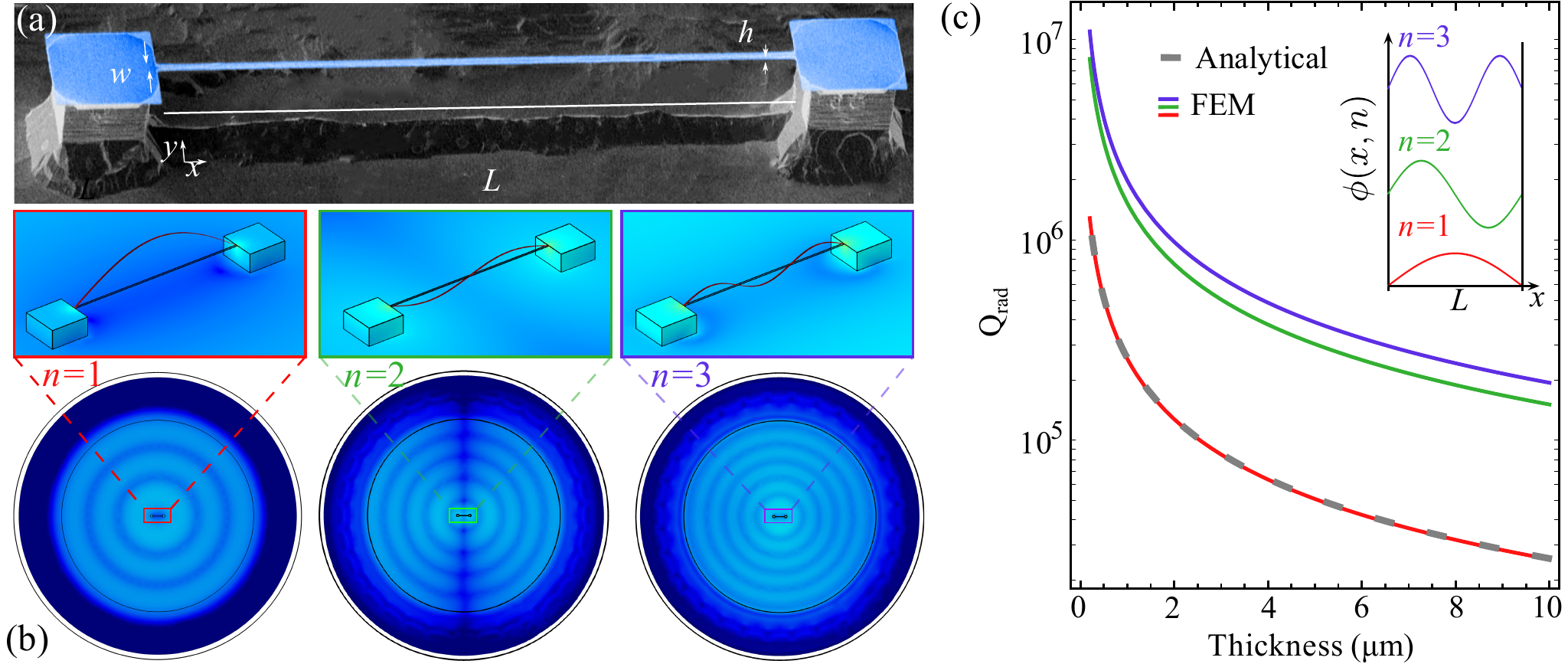}
\caption{(a) False color SEM image of a nanomechanical uniform string resonator on two pedestals, fabricated by the Quantum Optics Lab at the University of Queensland. The string resonator has a width of $w=5~\mu$m, thickness of $h=80$~nm and length of $L=1000~\mu$m. (b) Finite element modelling (FEM) simulations show the modeshape $\phi(x,n)$ of the first three eigenmodes ($n=1,2,3$) for the string resonator shown in (a). These simulations show the far field acoustic radiation pattern as acoustic waves transmit out of the resonator and propagate through the substrate. Note that for $n=1$ the far field resembles the far field of a monopole. For $n=2$ the two ends of the string are out-of-phase and the radiation pattern is a dipole. For $n=3$ the radiation pattern is a monopole with wavelength shorter than for the $n=1$ case. (c) Finite-element-method solvers modelling dissipation due to acoustic radiation for the first three eigenmodes $n=1,2,3$. The FEM simulations are compared with known analytical solutions for $n=1$. In this case, $Q_{\textnormal{rad}}$ of the fundamental mode is fit with the approximated analytical expression as a function of thickness, described in Ref.~\cite{romero_engineering_2020}, with the prefactor $\alpha=0.095$. Calculations correspond to silicon carbide uniform string of $L=1000~\mu$m and $w=5~\mu$m on a silicon substrate. } \label{fig:String}
\end{center}
\end{figure}

In the case of a stressed string resonator, clamping loss occurs at each of its rigid clamping points, as portrayed in Figure~\ref{fig:String}(b). The magnitude of this energy loss is determined by the ease at which acoustic energy stored in the string can propagate into the substrate. Due to the large size and mass difference between the string and its substrate, a large impedance mismatch exists at the connecting junction. The impedance mismatch causes a partial reflection of acoustic waves at the interface. The acoustic waves that do manage to travel through the junction carry some fraction of the total acoustic energy away from the resonator. Analytical approximations of radiation loss, and the quality factor associated with this loss $Q_{\textnormal{rad}}$ exist, and rely on finding the density of states overlap between the resonator and substrate modes \cite{cross_elastic_2001}. 

Previous works deriving clamping loss consider only a single clamped \cite{photiadis_attachment_2004} or unstressed string \cite{cross_elastic_2001}. An exact analytical solution for a stressed string has not been derived but $Q_{\textnormal{rad}}$ is known to scale with length and width of the resonator as $Q_{\textnormal{rad}} \propto L/w$~\cite{schmid_damping_2011,cross_elastic_2001}. Commonly the clamping loss of string resonators is calculated using finite element method (FEM) solvers \cite{ghadimi_radiation_2017,ghadimi_ultra-coherent_2018}. An advantage of using FEM solvers is the ability to alter the geometry of the clamping points and see its effect on $Q_{\textnormal{rad}}$ without needing to update or create new analytical models. In Figure~\ref{fig:String}(c) we find $Q_{\textnormal{rad}}$ based on finite element method solvers for the first three transverse modes and find the fundamental mode to agree well with the analytical approximation described in Ref.~\cite{romero_engineering_2020}. This analytical fit is a modification to the approximations described for a stressed membrane in Ref.~\cite{villanueva_evidence_2014}.

In general, maximizing the impedance mismatch between the substrate and resonator results in greater suppression of clamping loss. This has been demonstrated in trampoline resonators, in which increasing the substrate thickness, relative to the resonator thickness results in an increased $Q_{\textnormal{rad}}$~\cite{norte_mechanical_2016}. Similarly, since the impedance $Z$ of a material is given by $Z=\rho v_\phi$ where $v_\phi$ is the phase velocity~\cite{safavi-naeini_controlling_2019}, changing the material of either the resonator or substrate can help lessen the effect of clamping loss.

\subsubsection{Acoustic Bandgaps}

Phononic bandgaps arise from periodic arrays called phononic crystals (PnC), causing acoustic waves with carrier frequency within the bandgap to be unable to propagate. Mechanical modes within the bandgap experience a suppression of acoustic radiation into the substrate that is exponential with the number of unit cells in the crystal and can therefore be quite large~\cite{singh_high-q_2020,
tsaturyan_demonstration_2014,
yu_phononic_2014,ghadimi_radiation_2017,
tsaturyan_ultracoherent_2017}.
Bandgaps can be engineered to encompass the frequency of a desired mode by applying the correct periodicity $a$ of one repeated unit cell of the phononic crystal, much like a Bragg grating for the electromagnetic field~\cite{khelif_phononic_2016}. The width of the gap can also be engineered by careful choice of the geometry of the unit cell. 

\begin{figure}[h]
\begin{center}
\includegraphics[width=\textwidth]{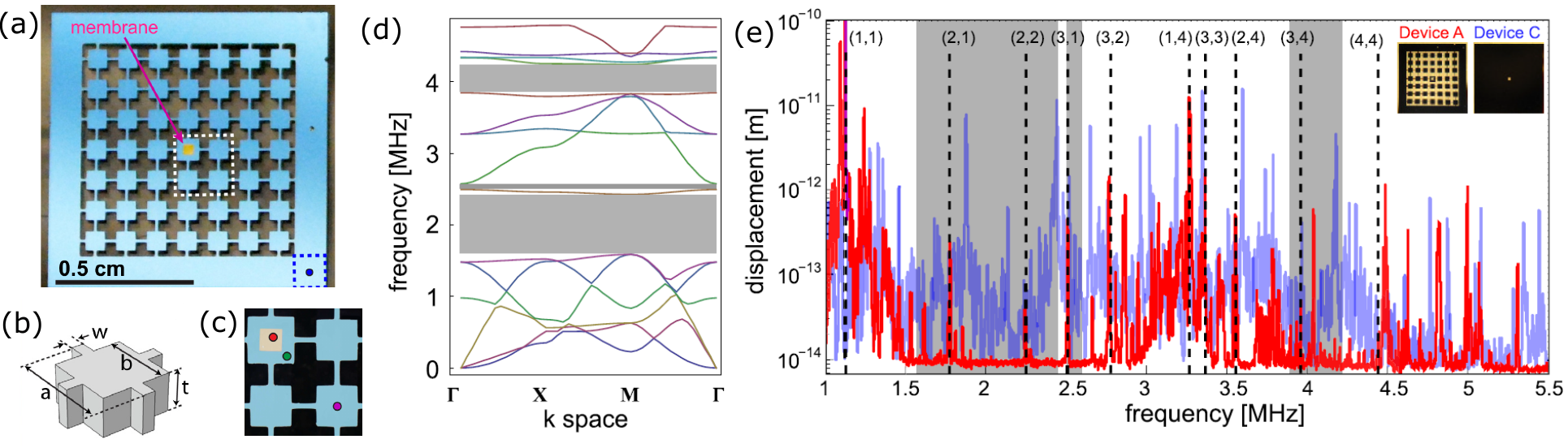}
\caption{(a) Optical image of a central squared Si$_3$N$_4$ membrane resonator (yellow) supported by a 2-D phononic crystal. (b) Unitary cell for of the phononic crystal, the geometric parameters of the cell $a$, $b$, $w$ and $t$ define the position width of the bandgap. (c) Zoom in of the central cell where the membrane has been fabricated. The red and purple dots denote the position of the light probe for the measured traces in (e). (d) Simulated dispersion relation for device A. (e) Measured mechanical displacement of the central membrane resonator mounted in a phononic crystal (red) and the phononic crystal frame (purple). Reproduced from P.-L. Yu $\emph{et al.}$ Applied Physics Letters $\textbf{104}$, 023510 (2014), with the permission of AIP Publishing.} \label{fig:Extrinsic_Losses2}
\end{center}
\end{figure}

Radiation losses in nanomechanical systems have been successfully suppressed by implementing a variety of phononic crystals. For example, Yu {\it et al.}~\cite{yu_phononic_2014} and Tsaturyan {\it et al.}~\cite{tsaturyan_demonstration_2014} simultaneously developed a two dimensional silicon phononic crystal that acoustically isolates a Si$_3$N$_4$ square membrane as shown in Figure~\ref{fig:Extrinsic_Losses2}(a). Here, acoustic waves generated by the motion of the membrane (red point in Figure~\ref{fig:Extrinsic_Losses2}(c)) are highly confined due to the phononic crystal, made of a squared lattice with a unitary cell shown in Figure~\ref{fig:Extrinsic_Losses2}(b). The phononic crystal opens a bandgap in the longitudinal, shear and transversal modes of the bulk silicon, as shown in Figure~\ref{fig:Extrinsic_Losses2}(d). The presence of the phononic bandgap results in up to three orders of magnitude of amplitude isolation from an external drive as shown in Figure~\ref{fig:Extrinsic_Losses2}(e), where the mechanical response for two different devices is measured. The mechanical responses presented correspond to a device directly mounted on the silicon substrate (Device C, purple trace) and a device supported by a phononic crystal (Device A, red trace). The comparison between these two measurements shows that within the bandgap, noise sources external to the membrane are highly suppressed. Several mechanical eigenmodes exist within the bandgap (dashed lines), and therefore have reduced clamping losses.

\begin{figure}[h]
\begin{center}
\includegraphics[width=0.9\textwidth]{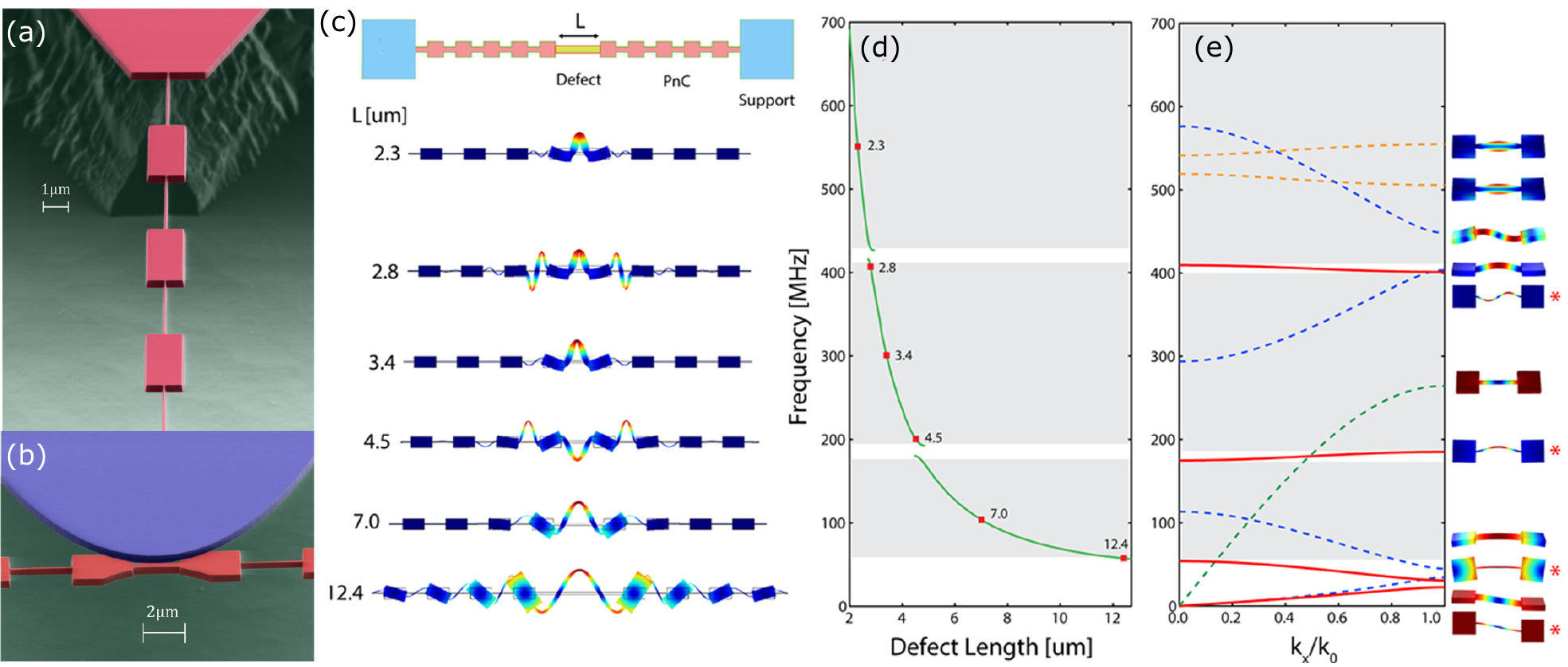}
\caption{(a) False color SEM image of a string resonator with periodic unit cells forming a phononic crystal (PnC). (b) Disc resonator (blue) used to couple and measure the displacement of PnC string, specifically with the center defect. (c) Schematic of the PnC string and FEM simulations of the eigemodes with defect length $L$ ranging from 2.3 $\mu$m to 12.4 $\mu$m. (d) Resonance frequency of the defect modes with length $L$ supported by a 1D phononic crystal. The red dots represent the eigenfrequency defined for the defect length shown in (c). (e) Dispersion relation for in-plane (red), out-of-plane (blue), breathing (green), and torsional modes (orange). Gray regions correspond to pseudo-bandgaps for in-plane modes. Adapted with permission from Ghadimi $\emph{et al}$. $\emph{Nano Letters}~\textbf{2017}$ 17(6), 3501-3505. Copyright 2017 American Chemical Society. } \label{fig:Extrinsic_Losses}
\end{center}
\end{figure}

 Phononic crystals can be implemented as a one dimensional array that isolates specific modes from propagating. For example, Ghadimi {\it et al.} demonstrated a 1D phononic crystal that selectively isolates in-plane flexural modes and confines them far away from the clamping points \cite{ghadimi_radiation_2017}. Figure~\ref{fig:Extrinsic_Losses}(a) shows a false color SEM image of their phononic crystal integrated on a string resonator. The phononic crystal is formed by unitary cell slabs that are connected longitudinally by a thin tether. At the center of the string, a defect cell is introduced, as shown in Figure~\ref{fig:Extrinsic_Losses}(b), where the string is coupled to a disk resonator for optical readout. The defect cell of length $L$ supported by the phononic crystal is represented in a schematic in Figure~\ref{fig:Extrinsic_Losses}(c). Here, the six in-plane flexural modes are plotted as a function of the defect length $L$. The six eigenfrequencies corresponding to these in-plane eigenmodes are displayed with red dots in Figure~\ref{fig:Extrinsic_Losses}(d). The phononic crystal is made by periodically modulating the width of the string resonator. The dispersion relations of the phononic crystal for different acoustic polarizations are shown in Figure~\ref{fig:Extrinsic_Losses}(e): in-plane (solid red), out-of-plane (dashed blue), breathing (dashed green), and torsional modes (dashed orange). The gray regions correspond to pseudo-bandgaps for in-plane modes.

As a final point on acoustic bandgaps, we would briefly note that an alternative method that has recently been developed to achieve such a bandgap, rather than creating a periodic repeating structure, is to utilise a zero-mode acoustic waveguide~\cite{mauranyapin_tunneling_2021}. In this case, no acoustic modes propagate at frequencies beneath a critical cut-off frequency. The dissipation of energy into the waveguide from mechanical resonances at lower frequencies is exponentially suppressed, similar to the exponential suppression from a periodic lattice.

\subsubsection{Nested Structures}\label{sec:Nested}

\begin{figure}[h]
\begin{center}
\includegraphics[width=.8\textwidth]{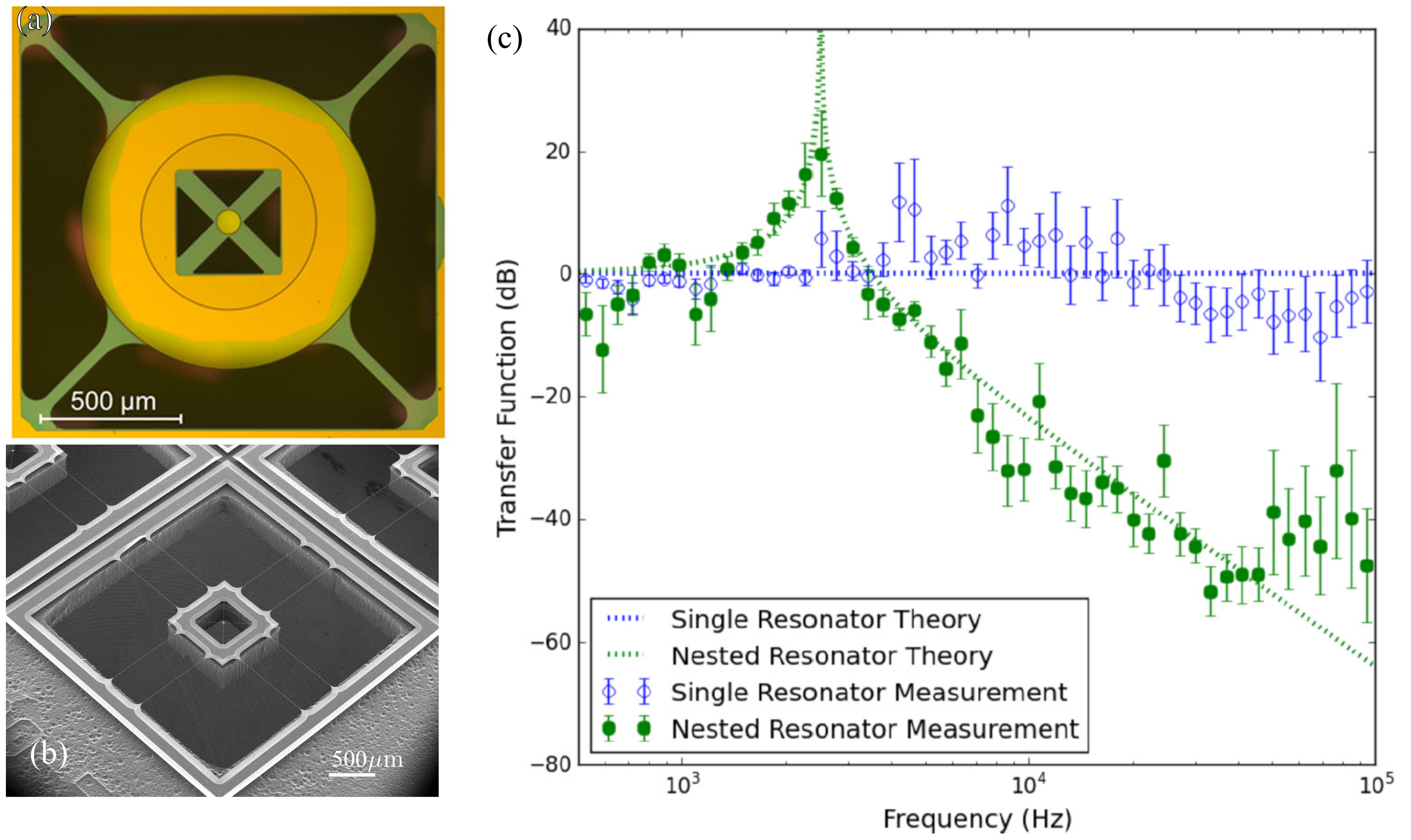}
\caption{(a) Optical image of a nested trampoline resonator from Ref.~\cite{weaver_nested_2016}. (b) SEM image of a nested resonator from Ref.~\cite{romero_sanchez_phononics_2019}. (c) Mechanical response of the device in (a). The chip was externally driven by a piezo and transmissions between the chip, outside the device, and the inner resonator were measured. This was done for both single and nested resonators. (a,c) Reproduced from M.J. Weaver $\emph{et al.}$ Nested trampoline resonators for optomechanics, Applied Physics Letters $\textbf{108}$, 033501 (2016), with the permission of AIP Publishing. (b) Reproduced from Ref.~\cite{romero_sanchez_phononics_2019} with permission from the author.}\label{fig:Nested}
\end{center}
\end{figure}

Nested structures can also be used for clamping loss suppression, emulating analog filters typically used in electronics. A well known macroscopic example of such an approach is the mass-spring stacks that have been implemented as acoustic filters for the vibration isolation of mirrors used in gravitational wave detection \cite{ballmer_new_2015,giaime_passive_1996,zhou_improved_1998}. In nanomechanics, nested low-pass filters have also proven to reduce the contribution of clamping loss. In this case, one resonator is nested within another ancillary resonator, or within a series of them~\cite{weaver_nested_2016,buters_where_2017,weaver_exploration_2018,romero_sanchez_phononics_2019}. Nested MEMS resonators have achieved a suppression power loss of $\sim$60 dB~\cite{buters_where_2017}, enhancing the radiative quality factor~\cite{weaver_nested_2016,romero_sanchez_phononics_2019}. An example of a nested MEMS resonator is shown in Figure~\ref{fig:Nested}(a), and its mechanical transmission as a function of frequency in Figure~\ref{fig:Nested}(c)~\cite{weaver_nested_2016}. The resonator consists of outer trampoline tethers, suspending a thick silicon mass and an inner nested trampoline resonator. It can be seen that the experimental data matches the theory predicted by the transfer function in Figure~\ref{fig:Nested}(c). In the simple case when there is only a single ancillary resonator, the power transfer function is
\begin{equation}\label{eq:transfer}
T(\omega_\textnormal{m}) =
\frac{\omega_\textnormal{a}^4}{(\omega^2_\textnormal{a}-\omega_\textnormal{m}^2)^2 +
\Gamma_\textnormal{a}^2\omega_\textnormal{m}^2}.
\end{equation}

This is a second-order low-pass filter, which has a slope of -40~dB/decade above the cut-off frequency. Here, $\omega_{\textnormal{m}}$ is the angular frequency of the isolated resonator, and $\omega_\textnormal{a}$ and $\Gamma_\textnormal{a}$ are, respectively, the angular frequency and damping rate of the ancillary resonator \cite{weaver_nested_2016,romero_sanchez_phononics_2019}. It is interesting to note that cascading $n$ nested resonators, in the form of mass-spring stacks in gravitational wave detection ancillary resonators emulates the characteristic $n$-th order transfer function of a Butterworth filter in electronics~\cite{butterworth_theory_1930}.

Indeed, once $\omega_\textnormal{m} \gg \omega_\textnormal{a}$ the suppression in amplitude for a single ancilla resonator scales as $(\omega_\textnormal{a}/\omega_\textnormal{m})^{2}$, compared for example to more abrupt filters made of four stacked resonator for which the scaling is $(\omega_\textnormal{a}/\omega_\textnormal{m})^{8}$~\cite{giaime_passive_1996}.

\section{Dissipation Dilution}\label{sec:Dissipation_Dilution}

Section~\ref{mech_sec} introduced the various forms of dissipation relevant to nanomechanical resonators, and showed that extrinsic loss mechanisms can be overcome by controlling the interaction between the nanomechanical resonator and its environment. In this section we explain the method of {\it dissipation dilution}, which can be used to improve the quality factor by lessening, or \emph{diluting} the intrinsic dissipation of a nanomechanical resonator.

\subsection{Origins of Dissipation Dilution} 

The concept of dissipation dilution first arose in the context of gravitational wave detection, where the exquisite precision required to detect faint sources necessitated the understanding and mitigation of intrinsic dissipation mechanisms \cite{gonzalez_brownian_1994,gonzalez_suspensions_2000}. 
The mirrors used in gravitational wave interferometers such as the Laser Interferometric Gravitational wave Observatory (LIGO) are suspended to isolate the measurements from acoustic and other vibrational noise sources. However, the act of suspending mirrors, while suppressing external noise sources, also introduces thermal noise to the detection. This noise scales linearly with the mechanical dissipation of the wires~\cite{schmid_fundamentals_2016,cleland_foundations_2003,gonzalez_suspensions_2000,PhysRevApplied.3.024004}. Consequently, achieving low wire dissipation is crucial to reach the ultimate limits of sensitivity. Through the optimization of suspended mirror designs, it was concluded that the dissipation can be reduced under a gravitational potential~\cite{gonzalez_suspensions_2000,gonzalez_brownian_1994}. This can be attributed to the resonator storing more energy under a gravitational potential, while leaving its mechanical damping mechanisms unaffected.
 Thus, the proportion of energy dissipated within one cycle is reduced and the resonator exhibits a longer decay time [see Equation~(\ref{eq:Q_FreqDissipation})]. 

Following LIGO's findings, the same phenomena of dissipation dilution was rediscovered in the context of nanomechanics when Verbridge {\it et al.} found that silicon nitride nanostrings under high-stress have higher than expected quality factors~\cite{verbridge_high_2006}. They had compared these silicon nitride strings to cantilever beams of the same frequency and found the strings to possess less dissipation. It was concluded that the enhanced quality factor and lower energy dissipation were a result of the high tensile stress. Since the motion of the string was normal to its high lateral tensile stress, it was able to store significantly more \emph{elastic potential energy} than a cantilever, similar to the addition of gravitational potential in the case of LIGO.
 
\subsection{Stress As a Lossless Potential}

Analogous to other resonators, the energy stored by a nanomechanical resonator oscillates between potential and kinetic energy. The potential energy $W$ of an unstressed resonator is stored in two forms, elongation  $W_{\textnormal{Elongation}}$ and bending $W_{\textnormal{Bending}}$. This potential energy is converted into kinetic energy during a nanomechanical resonator's oscillation cycle. This conversion between potential and kinetic energy is not without loss. As the resonator oscillates, the induced strain and bending give rise to all the intrinsic loss mechanisms described in Section~\ref{sec:IntrinsicLoss}. These can be divided into contributions from the resonator's elongation and bending. We define the energy lost per oscillation cycle due to each of these contributions as $\Delta W_{\textnormal{Elongation}}/2\pi$ and $\Delta W_{\textnormal{Bending}}/2\pi$.

Using the definition in Equation~(\ref{eq:Q_FreqDissipation}) the intrinsic quality factor of an unstressed resonator can then be expressed as 
\begin{equation}\label{eq:QintStoredELostCycle}
    Q_{\textnormal{int}}=2\pi \frac{{W_{\textnormal{Elongation}}}+{W_{\textnormal{Bending}}}}{\Delta W_{\textnormal{Elongation}}+ \Delta W_{\textnormal{Bending}}}.
\end{equation}

Applying tension to a nanomechanical resonator introduces a third contribution to the potential energy, $W_{\textnormal{Tensile}}$. Unlike the bending and elongation energies, which experience losses due to frictional forces, in the linear regime ($u_0 \gg h$) the tensile energy is lossless. Thus, the quality factor of a stressed resonator can be written as

\begin{equation}\label{eq:QD}
    Q_D=Q_{int}\times \left (1+ \frac{{W_{\textnormal{Tensile}}}}{ W_{\textnormal{Elongation}}+  W_{\textnormal{Bending}}} \right )= Q_{int} \times D.
    \end{equation}
The term $D$, usually referred to as the {\it dissipation dilution factor}, quantifies the fractional increase in mechanical quality factor due to the tensile stress.

\subsection{Quantifying the Dissipation Dilution Factor}

The tensile, elongation and bending energy of a nanomechancial resonator can be calculated from its material properties and resonant modeshape ($\phi_n(x)$ in the one-dimensional limit). Transversal modes are commonly targeted when engineering the dissipation of nanomechanical resonators because they exhibit higher dissipation dilution than in-plane, torsional and breathing modes \cite{fedorov_generalized_2019,fedorov_mechanical_2020}. Therefore we will limit ourselves to describing these modes. The resonator's transversal deformation in these modes can be described using its displacement field $u(x,y,z)$, which under the assumptions described earlier in Section~\ref{sec:funstring}, reduces to $u(x)=u_{0,n} \phi_n(x)$, where $u_{0,n}$ represents the amplitude of maximum displacement of the $n$-th mode.

In this uni-dimensional limit, the tensile, elongation and bending energy are given respectively by~\cite{schmid_fundamentals_2016,fedorov_supplementary_nodate}
\begin{equation}\label{eq:WTension}
    W_{\textnormal{Tensile}}=\frac{\sigma A}{2}\int_0^L \left (\frac{du(x)}{dx}\right)^2 dx,
\end{equation}
\begin{equation}
    W_{\textnormal{Elongation}}=\frac{EA}{8}\int_0^L \left(\frac{du(x)}{dx} \right )^4 dx \label{eq:Welong},
\end{equation}
and
\begin{equation}
    W_{\textnormal{Bending}}=\frac{EI_y}{2}\int_0^L \left(\frac{d ^2 u(x)}{dx^2} \right )^2 dx \label{eq:Wbending}.
\end{equation}
As can be seen, the tensile energy is largely determined by the stress $\sigma$\, but is also influenced by the modeshape~\cite{schmid_fundamentals_2016,fedorov_supplementary_nodate}. On the other hand, the stored elongation and bending energies are not directly dependent on the tensile stress $\sigma$. However, since the stress is partly responsible for determining the resonator's modeshape, as we will soon show, it does indirectly affect these two energies. It is worth noting that the tensile and elongation energies are both determined by the first derivative of the modeshape, whereas the bending energy is determined by the second derivative. Or in other words, unlike the tensile and elongation contributions, the bending energy depends on the curvature of the mode.

It can be seen from Equation~(\ref{eq:Welong}) that the elongation energy is quartic in the deflection of the resonator. As such, it is a nonlinear term, responsible for the Duffing nonlinearity at large amplitudes~\cite{sfendla_extreme_2020,catalini_modeling_2021}. By contrast, the bending and tensile energies are quadratic in deflection. They are responsible for the linear restoring force that causes the resonator to oscillate. Consequently, in the usual case where the dynamics of the resonator is well described as linear, we can expect the elongation energy to be insignificant compared to the other two energy components. Indeed, taking the example of a string resonator with a fundamental sinusoidal modeshape ($\phi(x)=\sin( \pi x/L)$) it is possible to show, using the moment of inertia defined earlier [see Equation~(\ref{eq:eigenfreq})], that ($W_{\textnormal{Bending}}/W_{\textnormal{Elongation}}) \approx 4h^2/9u_{0,1}^2$. Accordingly, $W_{\textnormal{Bending}}$ is greater than $W_{\textnormal{Elongation}}$ as long as the displacement field satisfies $u_{0,1}<2h/3$. That is to say, the elongation energy may be neglected so long as the amplitude of oscillation is significantly smaller than the thickness of the string. We make this assumption henceforth. 
The dissipation dilution factor can then be simplified to 
\begin{equation}\label{eq:DWTWB}
    D \approx 1+ \frac{{W_{\textnormal{Tensile}}}}{W_{\textnormal{Bending}}}.
\end{equation}
Thus for a linear resonator the key to maximising the dilution factor $D$ is to maximize the ratio of tensile energy ${W_{\textnormal{Tensile}}}$ to that stored in bending ${W_{\textnormal{Bending}}}$. 

Approximating a string resonator's modeshape using a sinusoidal function is common due to the simplicity and general intuition the generalization provides \cite{ghadimi_elastic_2018,schmid_damping_2011}. We did just that to motivate the validity of Equation~(\ref{eq:DWTWB}) for stressed resonators with linear dynamics. However,  a sinusoidal approximation does not perfectly describe the true resonant modeshape of a stressed string \cite{unterreithmeier_damping_2010,schmid_damping_2011}. The discrepancy lies in the region closest to the clamping points, where the resonator is firmly clamped. As a direct consequence of the boundary conditions imposed by these clamps, sharp curvatures arise in the resonator, making a sine function inaccurate in this region.
To account for these effects, a more general modeshape must be used. Ref.~\cite{schmid_fundamentals_2016} derives the modeshape
\begin{equation}
    \phi_{n}(x)=\sin(\beta_{\sigma} x) - \frac{\beta_{\sigma}}{\beta_{E}}(\cos(\beta_{\sigma} x) - e^{-\beta_E x}),
    \label{eq:correctedmodeshape}
\end{equation}
where $\beta_{\sigma}=n\pi/L$ represents the wavenumber and  $\beta_E=\sqrt{\sigma A/E I_y}$ accounts for the flexural rigidity. 

\begin{figure}[h]
\includegraphics[width=1\textwidth]{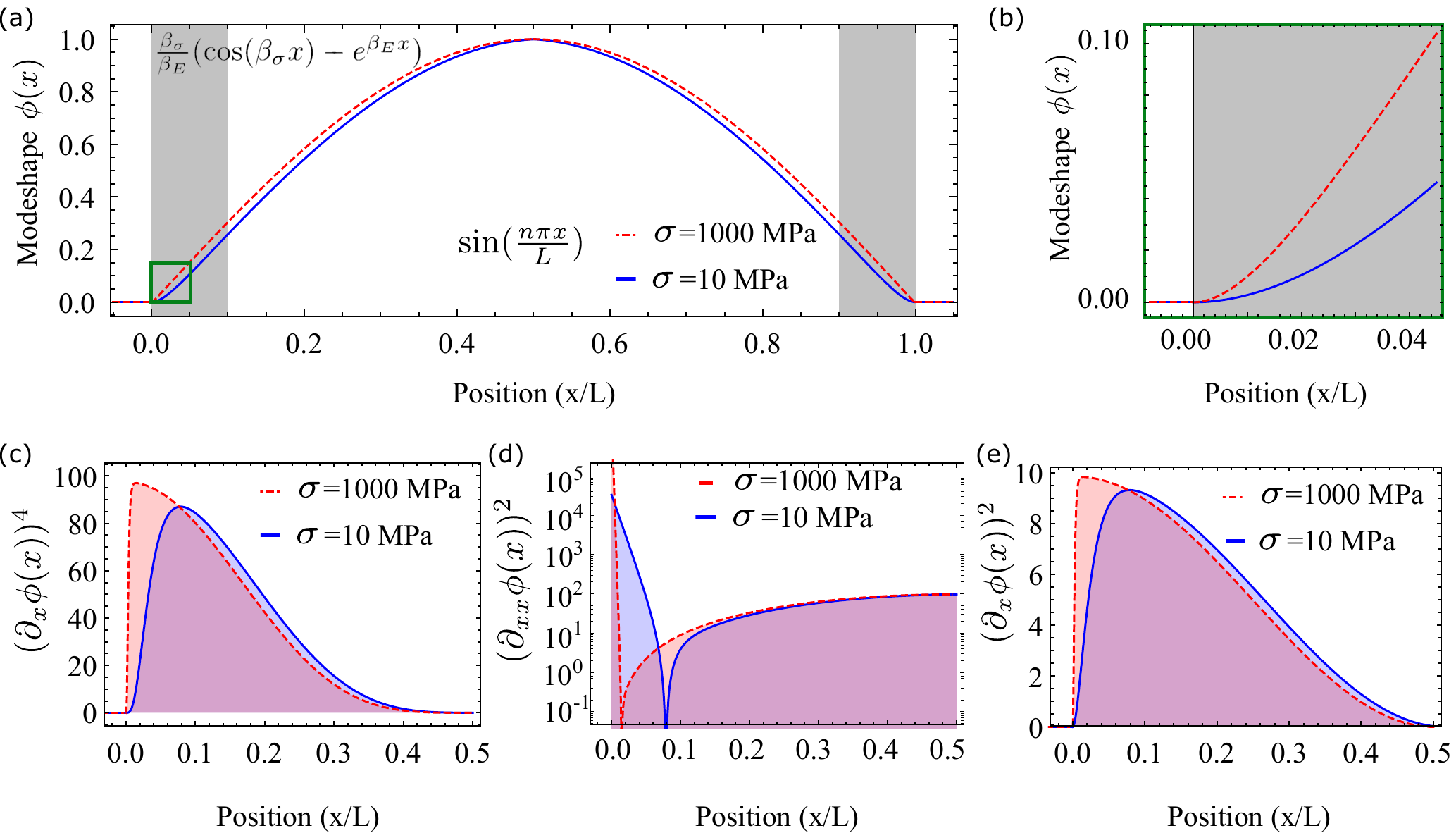}
\caption{(a) Modeshape $\phi(x/L)$ normalized to $x/L=1$ and $\phi(0.5)=1$ for the fundamental mode of a uniform string resonator with $\sigma=1000~$MPa (red) and $\sigma=10~$MPa (blue). (b) Zoom in of the modeshape shown in (a) for the two different stress regimes. (c) Geometrical energy density of $W_{\textnormal{Elongation}}$ in Equation \eqref{eq:Welong}. (d) Geometrical energy density of $W_{\textnormal{Bending}}$ in Equation \eqref{eq:Wbending}. (e) Geometrical energy density of $W_{\textnormal{Tensile}}$ in Equation \eqref{eq:WTension}.} 
\label{fig:Modeshape}
\end{figure}

 Figure~\ref{fig:Modeshape}(a) shows the fundamental mode of a stressed string with two different prestresses, using the more appropriate modeshape defined in Equation~(\ref{eq:correctedmodeshape}). Both modeshapes largely resemble a sinusoidal curve, but deviate near the clamping points [see Figure~\ref{fig:Modeshape}(a-b)]. Figures~\ref{fig:Modeshape}(c-e) plot the integrands within Equations~(\ref{eq:WTension},~\ref{eq:Welong},~\ref{eq:Wbending}) to illustrate the elongation, bending and tension energy densities.
 They show that the modeshape modification due to stress significantly changes the amount and location of stored energy. Indeed, Figure~\ref{fig:Modeshape}(d)  shows that increasing the stress from 10 to 1000~MPa concentrates the bending energy very near the clamping points and increases the energy density near them by two orders-of-magnitude.

\begin{figure}[h]
\includegraphics[width=1\textwidth]{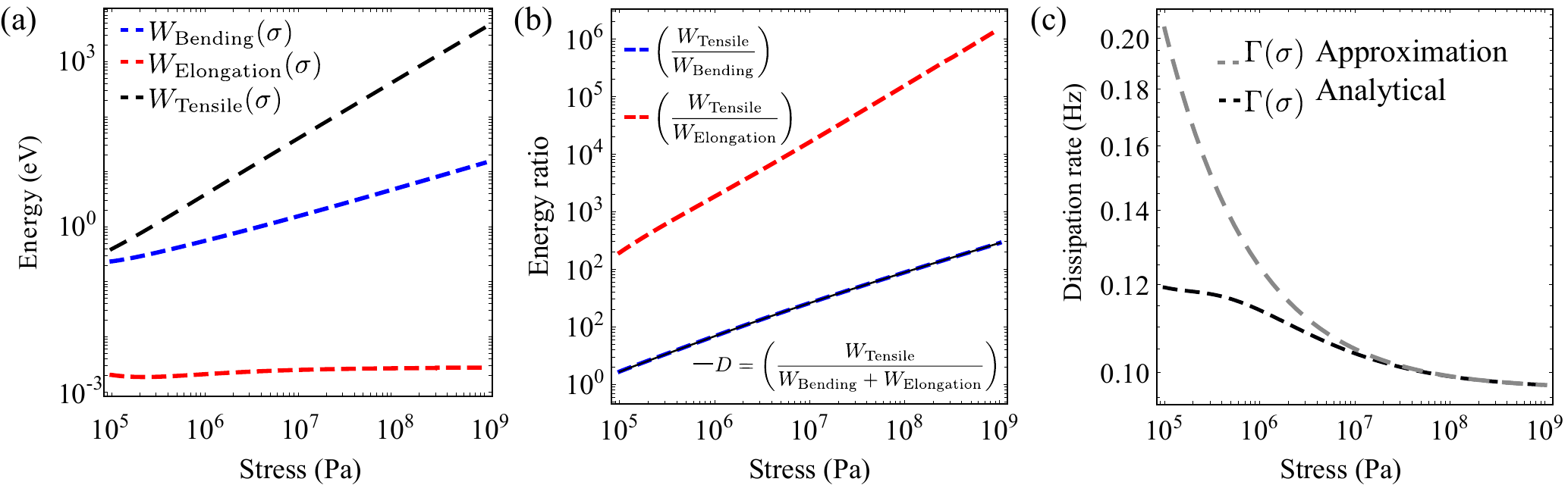}
\caption{(a) Distribution of potential energy stored as bending $W_{\textnormal{Bending}}$, elongation $W_{\textnormal{Elongation}}$ and tensile $W_{\textnormal{Tensile}}$ as a function of stress. (b) Energy ratio as a function of tensile stress between the tensile energy and the other two components of the elastic potential energy, bending and elongation. (c) Normalized dissipation rate as a function of stress. Found using the approximation made in Equation~(\ref{eq:correctedmodeshape}) (dashed light gray) and from full analytical solution including elongation (dashed black). These calculations were made for the fundamental mode of a uniform 3C-SiC string resonator with $L=1000~\mu$m, $w=1~\mu$m and $h=300~$nm. The calculations also consider $u_0=0.1~h$ for comparative reasons, specifically due to the elongation energy being many order of magnitude smaller at (more common) lower drive amplitudes.}
\label{fig:EnergyDistribution}
\end{figure}

For a more thorough comparison, we plot the energy stored in bending $W_{\textnormal{Bending}}$, elongation $W_{\textnormal{Elongation}}$ and tension $W_{\textnormal{Tensile}}$ as a function of tensile stress in Figure~\ref{fig:EnergyDistribution}(a) for a uniform string of typical size. It is evident that the modeshape $\phi(\sigma)$ is stress dependent since the bending $W_{\textnormal{Bending}}$ and elongation $W_{\textnormal{Elongation}}$ energies change as a function of stress while possessing no direct dependency [see Equations~(\ref{eq:Welong},~\ref{eq:Wbending})]. It is also apparent that the three forms of stored energy possess different scalings as a function of stress, indicating that stress can drastically alter the way a resonator distributes its stored energy. The elongation energy changed only marginally with increased stress, while the bending and tensile energies scale roughly to the square root and linearly with stress, respectively. Since the tensile energy increases fastest with stress, it is clear from Equation~(\ref{eq:DWTWB}) that increasing stress will promote dissipation dilution. We quantify this enhancement in dissipation dilution as a function of stress in Figure~\ref{fig:EnergyDistribution}(b), noting that at around 1~GPa a uniform string possesses a dilution factor on the order of hundreds for its fundamental mode. From this figure it is also clear that at typical oscillation amplitudes the elongation energy contribution to the dilution factor is negligible.

Neglecting the elongation energy (and therefore using Equation~(\ref{eq:DWTWB})) and using the corrected modeshape in Equation~(\ref{eq:correctedmodeshape}) to account for the sharp curvatures at the clamps we can simplify the dissipation dilution factor of a uniform string resonator to be \cite{schmid_fundamentals_2016,sadeghi_influence_2019}
\begin{equation}
    D \approx \Bigg[\frac{(n\pi)^2}{12}\frac{E}{\sigma}\Big(\frac{h}{L}\Big)^2 +\frac{1}{\sqrt{3}}\sqrt{\frac{E}{\sigma}}\Big(\frac{h}{L}\Big)\Bigg]^{-1} 
    \label{eq:DDApproxCorrect}.
\end{equation}
It is noticeable that a large aspect ratio $L/h$ is advantageous to increase the dissipation dilution factor, with $D$ scaling as $L/h$ in the high stress limit. This is particularly beneficial since the clamping loss also reduces as the device's aspect ratio increases \cite{villanueva_evidence_2014,norte_mechanical_2016}. It is also noticeable that $D\propto \sqrt{\sigma}$ at high stresses, as can also be seen in Figure~\ref{fig:EnergyDistribution}(b).
 
 The form of Equation~(\ref{eq:DDApproxCorrect}) allows the contribution to the dissipation dilution from the bending at the clamping points to be compared to the contribution from bending found along the remainder of the string \cite{schmid_fundamentals_2016,schmid_damping_2011}. The first term can be attributed to the portion of the string which largely resembles a sine function, while the second term represents the regions close to the clamping points shown in Figure~\ref{fig:Modeshape}(b). At levels of substantial stress, in our string example $\sigma > 10^5$ Pa, the second term representing the bending near the clamps dominates the dilution factor. Therefore it is the bending at the clamping points, rather than the entire length of the string, that is generally the limiting factor for the level of achievable dissipation dilution.

 So far in this section we have discussed the effect of dissipation dilution on the quality factor of a string resonator. However, the intrinsic quality factor depends on both the resonance frequency and intrinsic dissipation rate. While it is clear that increased stress will increase the resonance frequency ($\omega_\textnormal{m} \propto \sqrt{\sigma})$ [see Equation~(\ref{eq:eigenfreq})], the effect on the dissipation is less obvious and relatively unexplored in literature. To examine this, we show the change in the decay rate as a function of stress for a string resonator in Figure~\ref{fig:EnergyDistribution}(c). We plot this considering the approximation in Equation~(\ref{eq:DDApproxCorrect}) as well as the full analytical expression. The difference between the two lies in the fact that the latter includes the contribution of elongation energy towards the dilution factor. We see at high stress their differences are indiscernible, consistent with the rest of our assumptions and approximations. Both models predict a modest reduction in the dissipation with stress. Using the full analytical expression the reduction in dissipation asymptotes to a maximum of around 20~\% in the high stress limit.

\section{Soft Clamping}\label{sec:softclamp}

In Section~\ref{sec:Dissipation_Dilution} we showed that the bending energy is highly confined near the clamping points of a uniform string resonator, and that the contribution of this bending limits the maximum achievable dissipation dilution factor. Following this insight, one might anticipate that a lessening of bending, precisely in this region, could significantly improve the dissipation dilution factor. A technique highly effective in reducing bending at the resonator clamping points is known as \emph{soft clamping} \cite{tsaturyan_ultracoherent_2017,ghadimi_elastic_2018}.

\begin{figure}[h]
\includegraphics[width=\textwidth]{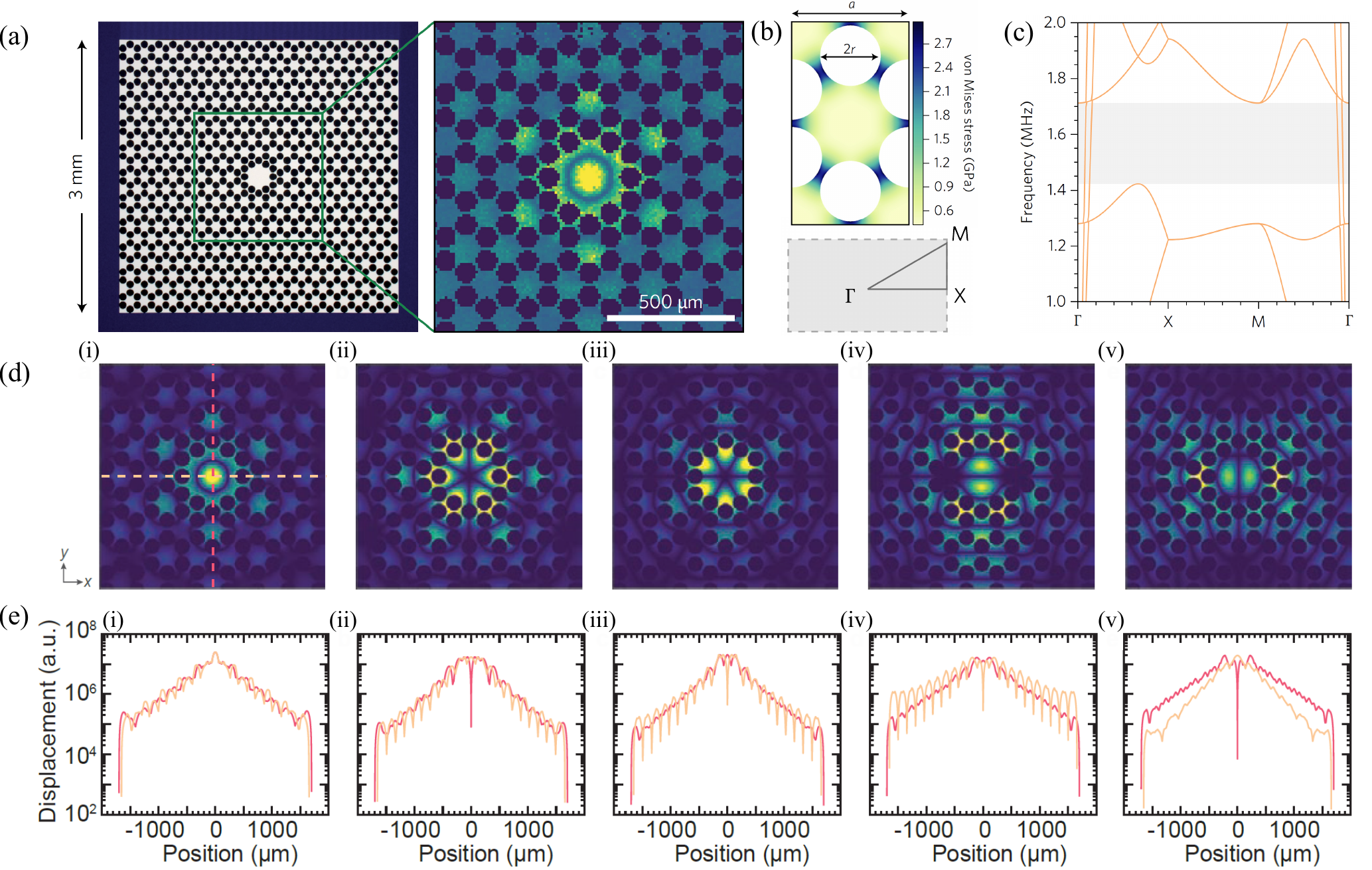}
\caption{(a) Micrograph of a patterned high stress silicon nitride membrane from Ref.\cite{tsaturyan_ultracoherent_2017}. The membrane consists of periodic hexagons and a centralized defect. Measured out of plane displacement for one localized mode. (b) Stress distribution of one rectangular unit cell. (c) Dispersion diagram, indicating a quasi-bandgap from 1.4MHz to 1.7MHz, highlighted in gray. (d) Simulations of many defect mode's amplitudes, all demonstrating some degree of soft clamping. (e) Corresponding simulated displacement fields, taken by two normal cuts shown in (di). Adapted by permission from Springer Nature: Nature Nanotechnology Y. Tsaturyan \emph{et al.} Ultracoherent nanomechanical resonators via soft clamping and dissipation dilution. $\emph{Nature Nanotechnology} ~ \textbf{12}$, 776-783(2017), Copyright 2017.   }\label{fig:Soft_Clamping_1}
\end{figure}

Soft clamping reduces bending by  altering the boundary conditions of the resonator~\cite{tsaturyan_ultracoherent_2017}, engineering its geometry to produce a gradual transition in the gradient of the modeshape between the clamping points and the rest of the resonator.
This can be interpreted as a progressive change in impedance along the resonator, to remove the large discontinuous impedance mismatch between the resonator and substrate. While a small impedance mismatch is beneficial in reducing a resonator's bending energy $W_{\textnormal{Bending}}$, it is unfavorable for clamping loss [see Section~\ref{sec:SuppressExtrinsicLoss}]. To overcome this trade-off, soft clamping techniques are commonly paired with the techniques to suppress clamping losses discussed in Section~\ref{sec:SuppressExtrinsicLoss}.

Simple changes to the clamp geometry have shown to be an effective way to enhance dissipation dilution. Small changes such as altering the width of the resonator near the clamping points have demonstrated some degree of soft clamping in string and trampoline resonators~\cite{sadeghi_influence_2019,romero_sanchez_phononics_2019}. For example, the rounding of a string resonator's clamps has shown to increase the quality factor by a factor of two by lessening the curvature of the mode~\cite{sadeghi_influence_2019,romero_sanchez_phononics_2019}. 

Another approach that enables soft clamping is the use of higher order defect modes in phononic crystals. These modes possess the highest amplitude in the center of the resonator away from the clamping points. This amplitude then gradually lessens as the mechanical wave hits additional potential barriers, which form the phononic crystal. With enough unit cells, the bending near the clamps can be greatly reduced. This has proven to be a very effective way to eliminate the storage of bending energy at a resonator's clamping points, thus increasing its dissipation dilution. For instance, it has enabled nearly an order of magnitude enhancement in the quality factor of $\textnormal{Si}_3\textnormal{N}_4$ string resonators \cite{ghadimi_radiation_2017}.

Soft clamping via higher order defect modes has similarly been incorporated into membranes \cite{tsaturyan_ultracoherent_2017,fedorov_thermal_2020}. Figure~\ref{fig:Soft_Clamping_1} shows a honeycomb membrane demonstrating several soft-clamped modes~\cite{tsaturyan_ultracoherent_2017}. The structure consists of a two-dimensional array of periodic and rectangular unitary cells show in Figure~\ref{fig:Soft_Clamping_1}(b), creating a phononic bandgap on the membrane modes, shown in Figure~\ref{fig:Soft_Clamping_1}(c). A defect mode is engineered to be localized at the center of the membrane, as shown in Figure~\ref{fig:Soft_Clamping_1}(a). The phononic bandgap confines additional mechanical modes mainly within the defect as shown in Figure~\ref{fig:Soft_Clamping_1}(d). Soft clamping is achieved by gradually modifying the rectangular lattice close to the defect, reducing the bending exhibited on the confined modes. The soft-clamping here, in a highly stressed $\textnormal{Si}_3\textnormal{N}_4$ material, dilutes the intrinsic loss by five orders of magnitude \cite{tsaturyan_ultracoherent_2017}.

Recently an alternative technique, specifically the use of hierarchical structures, has been proposed and experimentally realized to exhibit a large degree of soft clamping \cite{fedorov_fractal-like_2020,beccari_hierarchical_2021}. These resonators consist of cascading branches of beams, forming junctions under tension between the clamping points and the center of the resonator. At these junctions the overall curvature of the mode is lessened, compared to that of a single beam. The culminating effect of many branches collectively then results in extremely effective soft clamping, with virtually no bending occurring near the clamping points. Interestingly, a separate work using topology optimization arrives at a similar geometry~\cite{hoj_ultra-coherent_2021,gao_systematic_2020}. In this work, they use computer-aided geometrical optimization to determine an ideal resonator geometry. This optimization works to maximize the quality factor and frequency of a resonator's fundamental mode by minimizing its bending energy and radiation loss.

\section{Strain Engineering}\label{sec:strainengineer}

Soft-clamping is an effective way to increase a resonator's dissipation dilution by minimizing the elastic potential energy stored as bending. While this has led to orders of magnitude improvements in resonator quality factors~\cite{tsaturyan_ultracoherent_2017}, the form of the dilution factor in Equation~(\ref{eq:DWTWB}) suggests that further improvements should be possible by focusing on increasing the 
 tensile energy $W_\textnormal{Tensile}$.

As its name implies, the tensile energy is directly related to the tensile stress in a resonator [see Equation~(\ref{eq:WTension})]. Commonly stress is incorporated via the growth process of the film from which the resonator is fabricated. Deposition stresses, $\sigma_{\textnormal{dep}}$, of up to $\sim$~1~GPa are regularly achieved through these growth processes in $\textnormal{Si}_3\textnormal{N}_4$ and $\textnormal{SiC}$~\cite{reinhardt_ultralow-noise_2016,ghadimi_elastic_2018,kermany_microresonators_2014}. While considerable, this magnitude of stress is well below the \emph{yield strength} of these materials, which quantifies the maximum stress tolerated before fracturing. For example $\textnormal{Si}_3\textnormal{N}_4$ has a yield strength of 6~GPa~\cite{norte_mechanical_2016}, while it is  21~GPa for $\textnormal{SiC}$~\cite{romero_engineering_2020}, and 35~GPa for diamond~\cite{ruoff_yield_1979}. Furthermore, stress relaxation occurs upon release of nanomechanical resonators which means that, in practise, the stress of a uniform string resonator will always be below the stress of the unreleased film.

The relaxation of stress on release occurs due to the removal of the transverse stress in the resonator. This can be understood intuitively by considering a rubber band. As the rubber band is pulled longitudinally, its width and thickness compress. Analogously, after release the width and thickness of a nanomechanical resonator compress to relax some of the resonators longitudinal stress. The magnitude of this compression is determined by the Poisson ratio $\nu$ of the material. For example, a uniform string resonator will always relax to a stress of  $\sigma=\sigma_{\textnormal{dep}}\times (1-\nu)$, independent of its width~\cite{fedorov_mechanical_2020}.
Since the Poisson ratios of materials that are typically used in nanomechanics are in the range of $\nu\sim~$0.25, surpassing $\sim 75$\% of the deposition stress in string resonators is challenging~\cite{romero_sanchez_phononics_2019,ghadimi_ultra-coherent_2018}.

\subsection{Local Stress Enhancement}\label{sec:StressEnhancement}

The above discussion leads to the question of whether it is possible to achieve stresses beyond the deposition stress by the use of other means. Recent developments in semiconductor and nanomechanical systems have shown the feasibility of locally enhancing stresses by engineering a nanowire or resonator's geometry~\cite{norte_mechanical_2016,ghadimi_elastic_2018,minamisawa_top-down_2012}. In instances where this idea has been exploited, it has allowed for maximal local stresses of 3.8~GPa \cite{ghadimi_elastic_2018} and 6~GPa \cite{norte_mechanical_2016} in $\textnormal{Si}_3\textnormal{N}_4$ resonators, and 7.6~GPa \cite{minamisawa_top-down_2012} in Si nanowires, approaching the yield strength of both materials.

\begin{figure}[h]
\includegraphics[width=1\textwidth]{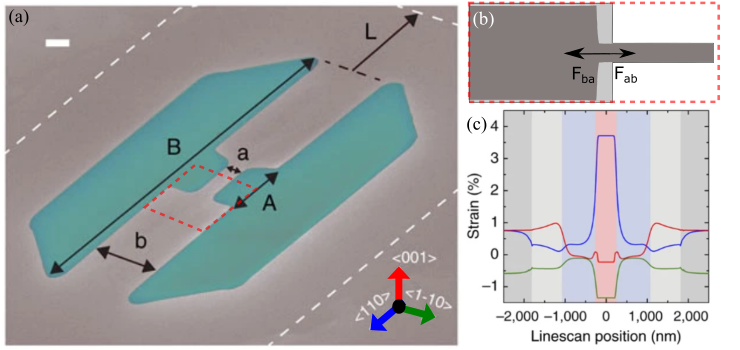}
\caption{(a) Colored SEM image of a suspended silicon bridge. The total length of the bridge is denoted B, with length A being the length of the narrow region. Their respective widths are denoted b and a. White scale bar in the top left corresponds to 300nm. (b) FEM simulation of one half of the device presented in (a). Here the light gray represents the geometry of the string before release, and black represents the structure in equilibrium after release and transfer of stress. The forces acting on the interface (black arrows), whose vector length corresponds to force magnitude. $F_{ba}$ represents the pulling force exerted by region $b$ on region $a$, and vice versa. (c) Simulated strain profiles along one cut along the length of the bridge. The blue trace corresponds to the strain profile along the beam in the blue or $<110>$ direction in (a). The red and green traces follow the same cut along the length but represent the strain profiles in the $<001>$ and $<1-10>$ directions respectively \cite{minamisawa_top-down_2012}. Adapted by permission from Springer Nature: Nature Communications, Minamisawa, R. $\emph{et al}$.  Top-down fabricated silicon nanowires under tensile elastic strain up to 4.5$\%$.$\emph{ Nat Commun} ~ \textbf{3}$, 1096 (2012), Copyright 2012.  }
\label{fig:StressRelax_SiNanowire}
\end{figure}

To motivate how choice of geometry can create local stress maxima we consider the case of a prestressed string with width that is nonuniform, varying along its length, such as that presented in Figure~\ref{fig:StressRelax_SiNanowire}(a) \cite{minamisawa_top-down_2012}. For simplicity, and since resonators are typically fabricated by under-etching a uniform thin film, we consider the thickness $h$ to be constant. We also assume that the total length of the resonator remains fixed, since it is rigidly clamped in this direction. Along with the stress relaxation exhibited by a uniform string, when released from its substrate, a nonuniform string also exhibits stress transfer. This is required to maintain a constant force of tension $F_T$ along the string. One half of the string presented in Figure~\ref{fig:StressRelax_SiNanowire}(a) can be seen in Figure~\ref{fig:StressRelax_SiNanowire}(b). Here, at the interface between the wider and narrower regions, we denote the two acting tension forces. These are the tension forces of region $a$ acting on region $b$, $F_{ab}$, and vice versa $F_{ba}$. Since the tension force is equal to $F_T=\sigma A$, where $A$ is the resonator's cross sectional area, the total force $F_{\textnormal{Total}}$ at the interface immediately after release of the structure is $F_{\textnormal{Total}}=A_a \sigma_{a}-A_b \sigma_b$. For a deposition prestress $\sigma_{\textnormal{dep}}$, $F_{\textnormal{Total}}=(A_a-A_b) \sigma_{\textnormal{dep}}(1-\nu)$. Designing the areas of regions $a$ and $b$ to be different, means at the moment of release of the structure there is a nonzero net force. As a result region $b$ pulls on region $a$ until the stress in region $a$ is elevated enough to make the net force zero. In this way, nonuniform geometries can allow local stresses that exceed the maximum deposition stress.

The amount of local stress enhancement relative to the deposition stress is determined by the width and length ratios of the narrow to wider regions. As a general rule, to increase the stress of the narrow region by a factor $C$ above the deposition stress, the wider regions need to be $C$ times wider and longer. This can be seen in the blue trace in Figure~\ref{fig:StressRelax_SiNanowire}(c), representing the strain profile $\epsilon$ (related to the stress by $\epsilon=\sigma/E$) along the string's length. Here the center region was chosen to be roughly 1.5 times shorter and 12 times narrower than the larger region, allowing for local strain enhancement of four times the deposition strain. It can also be seen that this local strain enhancement results in a decrease in strain everywhere else in the string. For more thorough analytical results, refer to \cite{fedorov_mechanical_2020,minamisawa_top-down_2012}.

\begin{figure}[h]
\begin{center}
\includegraphics[width=1\textwidth]{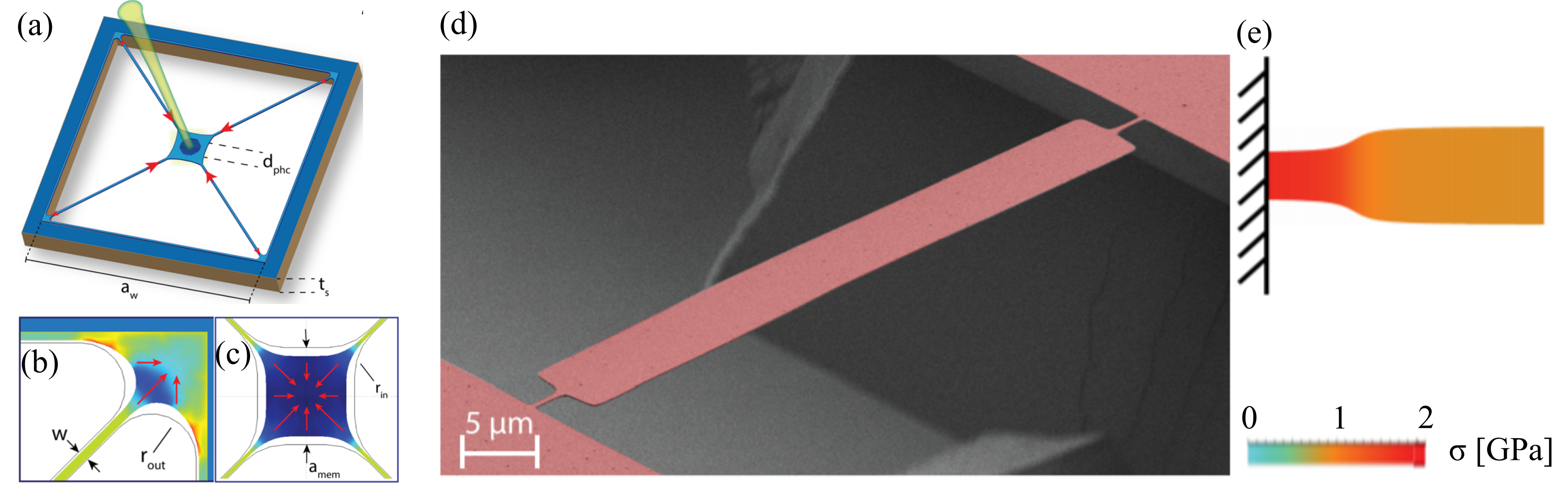}
\caption{(a-c) FEM simulations modelling the stress relaxation of trampoline resonators presented in \cite{norte_mechanical_2016}. (b-c) Stress distribution at the resonator's clamping points and central pad respectively, where the arrows denote length contraction. Reprinted with permission from R.A. Norte $\emph{et al.}$ PRL, $\textbf{116}$, 147202 (2016). Copyright 2016 by the American Physical Society. (d) False colored SEM image of clamp-tapered  string resonators from Ref.~\cite{bereyhi_clamp-tapering_2019}. (e) Schematic of relaxed stress distribution at one clamping point of the same devices in (d). Adapted with permission from Bereyhi \emph{et al}. $\emph{Nano Letters}~ \textbf{2019}$ 19(4), 2329-2333. Copyright 2019 American Chemical Society. }
\label{fig:Strain_Enhancement}
\end{center}
\end{figure}

The idea of local strain enhancement has been demonstrated with success in several nanomechanical devices \cite{norte_mechanical_2016,bereyhi_clamp-tapering_2019}, as shown for example in Figure~\ref{fig:Strain_Enhancement}. The trampoline resonators in Figure~\ref{fig:Strain_Enhancement}(a-c) achieved near-yield-strength stress as the larger width of the clamping points and center pad compared to the tethers allows for strain transfer to the tethers. This allowed quality factors as high as $Q \approx 10^8$ to be obtained, corresponding to dissipation rates of $\Gamma \approx 1.4 \textnormal{mHz}$ at room temperature \cite{norte_mechanical_2016}. Applying the same idea of varying the cross-section, but instead tapering near the clamping points of a string resonator has also been found to improve the quality factor, in this case by roughly a factor of two~\cite{bereyhi_clamp-tapering_2019}.

\subsection{Strain Engineering}  
While Refs.\cite{norte_mechanical_2016,bereyhi_clamp-tapering_2019} discussed in the previous section both achieve some level of dissipation dilution by engineering the local strain in nanomechanical resonators, the devices would at some point be limited because, although the maximal tensile stress of the resonator is increased, the wider regions of the resonator actually lose tensile stress. This essentially counteracts any improvements in dissipation dilution if the mode of the nanomechanical resonator extends into the wider region [see Equation~\ref{eq:WTension}]. Since it is common for the mode of a resonator to extend over the full length of the device, a localized stress enhancement does not necessary guarantee higher dissipation dilution. For example, consider the devices presented in Figure~\ref{fig:Strain_Enhancement}(a-c). The tethers, which cover a large portion of the mode, also possess enhanced stress through stress transfer from the clamping points and central pad. However, the modeshape extends over the full device, including both the central pad and clamping points which, due to stress transfer, have reduced local stress. So while the devices very likely benefited from local stress enhancement, better overlap of the locally elevated stress with the resonant modeshape could have allowed for greater dissipation dilution (assuming the dissipation was not clamping loss limited).

\begin{figure}[h]
\includegraphics[width=1\textwidth]{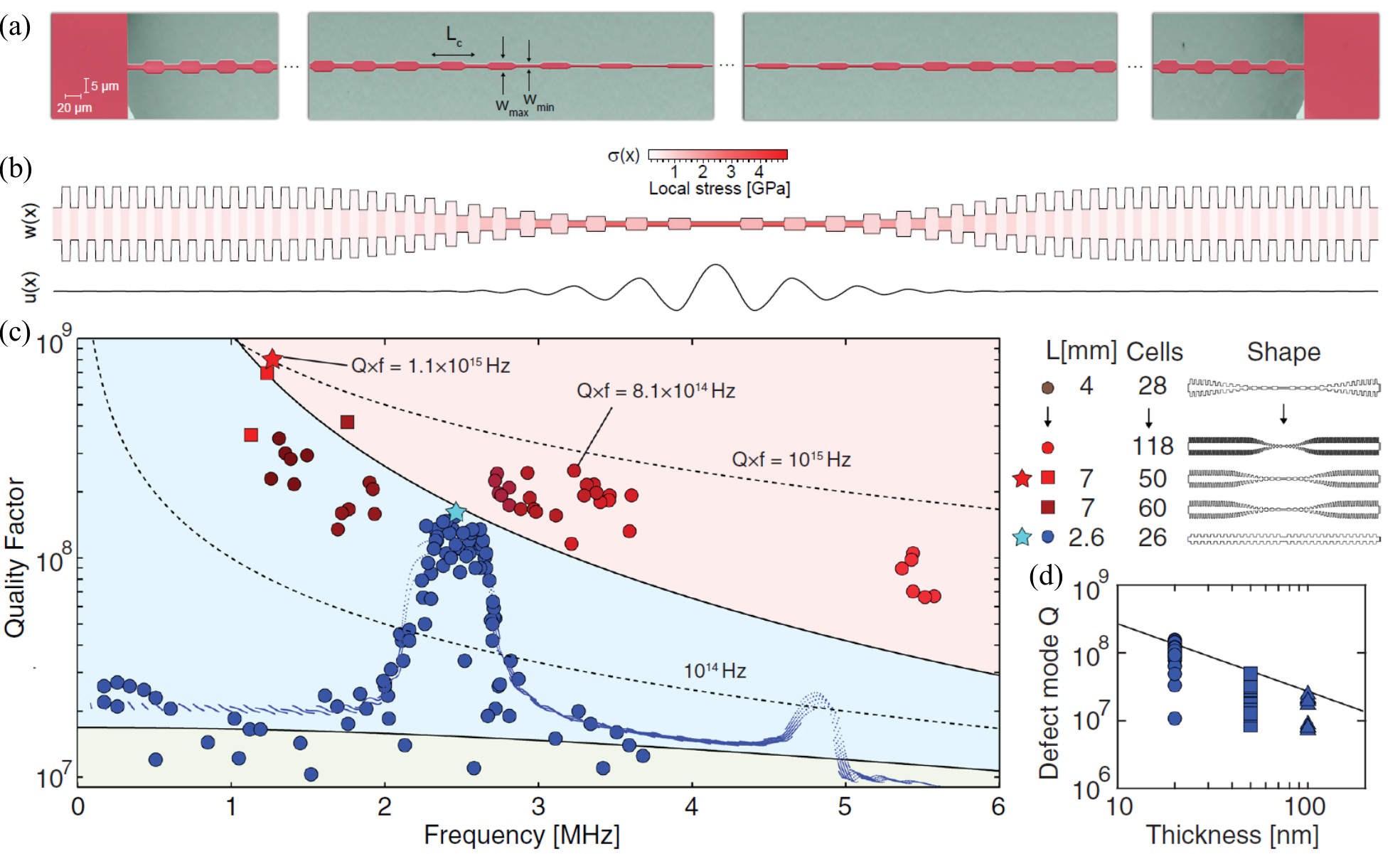}
\caption{(a) SEM image of a tapered non-uniform string resonator. (b) Width $w(x)$ and stress $\sigma(x)$ profiles of their resonators, as well as a high quality factor defect mode shape $u(x)$. (c) Q versus mode frequency for several geometries of strings, each geometry denoted by color to its right. (d) Defect mode quality factor versus thickness plot. From A.H. Ghadimi $\emph{et al., Science} ~ \textbf{360}$, 764-768 (2018). Reprinted with permission from AAAS.} 
\label{fig:Strain_Engineering}
\end{figure}

 We see from the above discussion that to fully benefit from local stress enhancement it is necessary to co-locate the mode of the resonator with the region of high stress. This technique is known as  \emph{strain engineering}, and was first demonstrated by  Ghadimi {\it et al.}~\cite{ghadimi_elastic_2018}. They used a tapered string cross-section to incorporate locally enhanced stress into a phononic crystal defect mode [see Figure~\ref{fig:Strain_Engineering}(a)], much like those presented in Section~\ref{sec:softclamp} \cite{ghadimi_elastic_2018}. They designed the structure of the resonator such that the nodes of one particular resonant mode were co-located spatially with the regions of lower stress, while the anti-nodes were co-located with regions of higher stress. In this way, the stress of the regions that exhibit large displacements is, on average, increased. This results in more stored tensile energy and a higher dissipation dilution factor.
 
Figure~\ref{fig:Strain_Engineering}(c) quantifies the enhancement gained  from incorporating strain engineering by comparing a tapered soft-clamped beam (red points) to a soft-clamped beam (blue points) \cite{ghadimi_elastic_2018}. It is evident that tapering the beam allows the resonator to achieve higher quality factors through additional dissipation dilution. This is unattainable with the untapered beams because, although soft-clamped, local stresses were not increased above the material's deposition stress. In other words, the difference between the quality factors can be attributed to the fact that one incorporates soft clamping only (blue points), while the other overlays strain with the regions of the displacement field in a soft clamped mode to accomplish strain engineering (red points). 

The devices presented in Figure~\ref{fig:Strain_Engineering}(a-b) showed a quality factor enhancement of up to an order of magnitude compared to untapered soft-clamped beams of similar frequencies. They demonstrate strain-engineering is a practical technique which can elevate deposition stress in a resonator to close to the material's yield strength. Without strain engineering, this would require higher deposition stress or external forces that enhance stress over the entire length of the resonator.

\subsection{Prospects for Strain Engineering}

\begin{figure}[h]
\includegraphics[width=\textwidth]{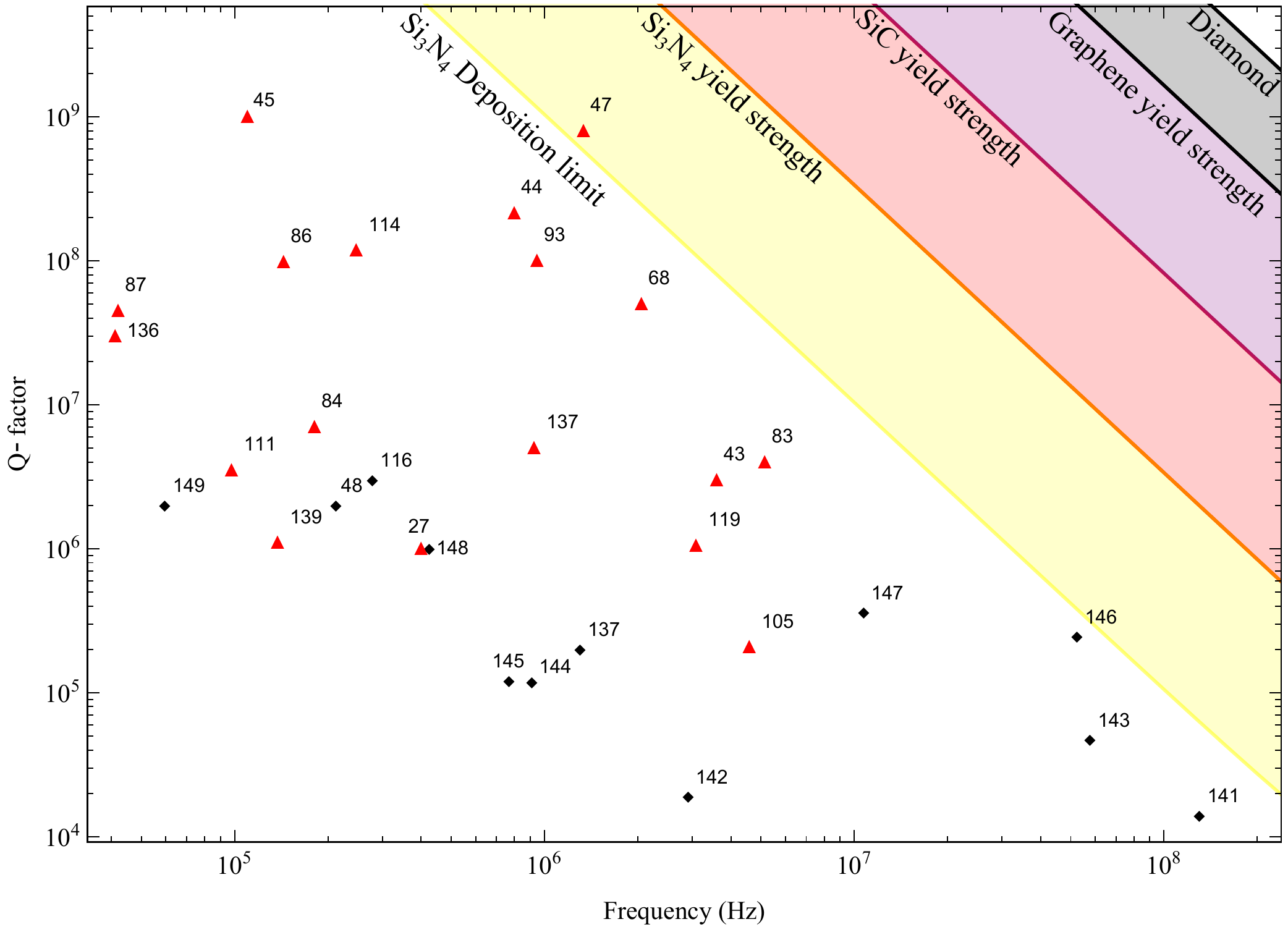}
\caption{Comparison of stressed silicon nitride (red) resonators to stressed alternative material resonators (black). Alternate materials include silicon carbide (3C-SiC), silicon nitride with aluminium, aluminum, gallium arsenide, gallium nitride,  graphene, tungsten diselenide, and indium gallium phosphide. The plot considers highly stressed nano- and micro-mechanical string, drum, trampoline and membrane resonators, some of which are at room temperature, others at cryogenic temperatures. Additional references incorporated in the figure, but not in the text include \cite{chakram_dissipation_2014,pluchar_towards_2020,yu_control_2012,verbridge_megahertz_2008,zwickl_high_2008} and  \cite{lee_high_2013,chen_performance_2009,yamaguchi_improved_2008,morell_high_2016,sang_strain-enhanced_nodate,onomitsu_ultrahigh-q_2013,will_high_2017,teufel_circuit_2011,cole_tensile-strained_2014,liu_high-q_2011} for alternative materials.} \label{fig:PlotsMap}
\end{figure}

Strain engineering has proven to be a powerful technique to enhance the dissipation dilution in nanomechanical resonators. The effectiveness of this technique is contingent on the use of a film which is intrinsically prestressed and has a high yield strength. For this reason, silicon nitride has been largely used to make highly diluted resonators. Highly stressed silicon nitride thin films are commercially available, and can be deposited on silicon, meaning they are compatible with well known nanofabrication procedures and equipment. Indeed, in general, stressed resonators fabricated from $\textnormal{Si}_3\textnormal{N}_4$ (red triangles)  have a record of allowing significantly higher quality factors than those fabricated from other materials (black diamonds), as shown in  Figure~\ref{fig:PlotsMap}. However, the local stresses engineered into these resonators now approach the yield strength of $\textnormal{Si}_3\textnormal{N}_4$, limiting the prospects for further improvements in quality factor.

To examine whether alternative materials could allow for better performing resonators in the future, we consider silicon carbide~\cite{kermany2014microresonators, romero_engineering_2020, hamelin_monocrystalline_2019}, graphene~\cite{PhysRevApplied.3.024004,muruganathan_graphene_2018,singh_giant_2020} and diamond~\cite{tao_single-crystal_2014,ovartchaiyapong_high_2012,najar_microcrystalline_2013}, comparing their upper-bounds of performance relative to silicon nitride. The upper-bounds are calculated by assuming the material's intrinsic quality factor is the highest experimentally demonstrated to our knowledge at room temperature~\cite{tao_single-crystal_2014,PhysRevApplied.3.024004,hamelin_monocrystalline_2019}, and their dissipation dilution limit is set by the material yield stress~\cite{ruoff_yield_1979,lee_measurement_2008,romero_engineering_2020}, as described in Ref.~\cite{fedorov_generalized_2019}. Compared to silicon nitride, all three proposed alternatives have higher yield strength limits, while at the same time silicon carbide and diamond have less intrinsic dissipation.

It can be seen from the shaded regions in Figure~\ref{fig:PlotsMap} that these alternative materials indeed enable significantly improved quality factors due to their favorable mechanical properties. Resonators made of highly stressed silicon carbide for example, allow for the possibility of quality factors over an order of magnitude higher than what is possible with silicon nitride. Graphene and diamond possess even more potential advantage, offering almost three and four orders of magnitude respectively in quality factor enhancement for a given resonance frequency. These improved quality factors could be instrumental in advancing many existing applications, such as sensing and filtering~\cite{sun_znosilicon_2012}, and also to realizing room temperature nanomechanical quantum technologies~\cite{aspelmeyer_cavity_2014,norte_mechanical_2016}. Additional advantages of alternative materials, specifically diamond and silicon carbide with low intrinsic dissipation, is that this could immediately translate into better performance using existing nanomechanical resonator geometries. For example, fabricating the $\textnormal{Si}_3\textnormal{N}_4$ soft-clamped devices from Ref.~\cite{beccari_hierarchical_2021} instead from diamond could reduce intrinsic dissipation by over an order of magnitude. This would translate into force sensitivities on the order of $\approx 10\textnormal{zN}~\textnormal{Hz}^{-1/2}$ at room temperature. Similar sensitivity gains could be seen across various sensing applications including mass and single-molecule sensing with the introduction of new materials with less intrinsic dissipation.

In addition to providing immediate improvement in nanomechanical performance, a shift towards materials like diamond and silicon carbide could create new research directions. For instance, in addition to their excellent mechanical properties, diamond and silicon carbide have proven to be ideal for integrated quantum photonics and solid-state qubits~\cite{lukin_4h-silicon-carbide--insulator_2019,lukin_silicon_2021,janitz_cavity_2020,nguyen_quantum_2019}. Therefore, the realization of ultra-low dissipation nanomechanical resonators made of diamond or silicon carbide could allow for the natural interfacing of nanomechanics and color centers \cite{castelletto_silicon_2020,lee_strain_2016,wang_coupling_2020,ovartchaiyapong_dynamic_2014}. This could allow for the realization of hybrid solid-state quantum systems \cite{poshakinskiy_optically_2019}.

\section{Outlook}

The utilization of dissipation dilution promises to significantly extend the applications and performance of nanomechanical resonators. Alongside techniques to control extrinsic dissipation, dissipation dilution has already allowed for over five orders of magnitude enhancement in a nanomechanical resonator's quality factor \cite{ghadimi_elastic_2018,tsaturyan_ultracoherent_2017,beccari_hierarchical_2021}. Further increases in dissipation dilution will likely occur through the use of new resonator geometries. As we have shown dissipation dilution is highly dependent on geometry, with recent works using topology optimization and hierarchical structures showing that new device geometries with better performance are still being found \cite{beccari_hierarchical_2021,hoj_ultra-coherent_2021}. It is likely that refined optimization techniques, and/or human insight will discover alternative geometries with better performance or geometries better suited for certain applications.

A shift towards other materials will certainly further enhance the quality factors that are accessible in nanomechanical resonators. Materials with low intrinsic dissipation and large yield strengths, such as the crystalline alternatives presented in Figure~\ref{fig:PlotsMap}, are ideal candidates for ultra-coherent resonators. While these alternative materials have high potential, experimental progress has been slowed by challenges in fabrication. As a result of their strength and lattice structure, fabrication and film growth of crystalline mechanical resonators presents real challenges. For example, in the growth of crystalline materials even slight defects can create propagating cracks throughout the material, and cause additional material loss \cite{romero_engineering_2020}. Despite these challenges, the massive yield strength of these materials, particularly graphene with a yield strength of 130~GPa~\cite{lee_measurement_2008}, offer great opportunities of dissipation dilution and strain engineering techniques.

\section{Acknowledgements}

This research is funded by the Australian Research
Council and Lockheed Martin Corporation through
the Australian Research Council Linkage Grant No. LP190101159. This research is partially supported by
the Commonwealth of Australia as represented by the
Defence Science and Technology Group of the Department
of Defence. Support is also provided by the Australian
Research Council Centre of Research Excellence for Engineered
Quantum Systems (Grant No. CE170100009).  This work was supported by the
Australian National Fabrication Facility, a company established under the National Collaborative Research Infrastructure
Strategy to provide nano- and microfabrication
facilities for Australia’s researchers. 

\bibliographystyle{MSP}


%

\end{document}